\documentclass[fleqn,12pt,twoside]{article}
\usepackage{espcrc1}

\usepackage{graphicx}
\usepackage{epsfig}

\usepackage[figuresright]{rotating}

\newcommand{\AmS}{{\protect\the\textfont2
  A\kern-.1667em\lower.5ex\hbox{M}\kern-.125emS}}

\title{Inclusive electron scattering off $^4$He }

\author{S. Bacca\address[GSI]{Gesellschaft f\"ur Schwerionenforschung, Planckstr.~1, 64291
Darmstadt, Germany}%
        \thanks{ Partially supported by the Deutsche
          Forschungsgemeinschaft/SFB 443.},
         H. Arenh{\"o}vel\address{Institut f\"ur Kernphysik, Johannes
           Gutenberg-Universit\"at,\\ Becher-Weg 45, 55099
Mainz, Germany},  N. Barnea\address{Racah Institute of Physics,
Hebrew University, 91904 Jerusalem, Israel},
W. Leidemann\address[TN]{Dipartimento di Fisica, Universit\`a di
  Trento and \\Istituto Nazionale di Fisica Nucleare, Gruppo
Collegato di Trento,\\ I-38050 Povo, Italy}
        and  
        G. Orlandini \addressmark[TN]
}
       
\begin{document}

\maketitle

\begin{abstract}
Inclusive electron scattering off $^4$He is calculated exactly
with a complete treatment of the final state interaction within a
simple semirealistic potential model. We discuss results for both the longitudinal and the
transverse response functions, at various momentum
transfers. A consistent meson exchange current is implemented.
Good agreement with available experimental data is found  for the longitudinal
response function, while some strength is still missing in the
transverse response function.
\end{abstract}

\section{INTRODUCTION}
Inclusive electron scattering is governed by two response
functions: the
longitudinal $R_L(\omega,{\bf q})$ and the transverse response
$R_T(\omega,{\bf q})$. They are induced by the electromagnetic charge $\hat{\rho}({\bf q})$ and  current $\hat{\bf J}({\bf q})$ operators, respectively.
We study this process on the nucleus of $^4$He, 
for which  an  exact  calculation of the 
response function can be performed,   including a consistent treatment of the
electromagnetic excitation operator.
The final state interaction of the continuum four-body wave function is fully taken into
account via the  Lorentz Integral Transform (LIT) method \cite{LIT}, which
leads to  a  Schr\"{o}dinger-like equation with bound-state-like asymptotic.
We solve it by  making   use of a
 spectral resolution method based on  the construction of an 
 effective interaction  in the hyperspherical harmonics  basis (EIHH) \cite{EIHH}.
For the present calculation we take the simple semirealistic
Malfliet-Tjon (MTI-III) \cite{MT} as nucleon-nucleon (NN) interaction. 

In a non-relativistic approach the  electromagnetic charge  is given by   a
one-body operator, while  the current by both a one-body and  a
two-body operator, the meson exchange current (MEC). 
We firstly show our calculation of the longitudinal response function
and then we present our result for the transverse response function,
where we consider also a consistent two-body current. 

\section{LONGITUDINAL RESPONSE FUNCTION}

The longitudinal response function is defined as
\begin{equation}
R_L(\omega,{\bf q} )= \int \!\!\!\!\!\!\!
\sum_f\left| \left\langle \Psi
    _{f}\right| \hat{\rho}({\bf q})\left| \Psi
    _{0}\right\rangle \right| ^{2}\delta
\left(E_{f}-E_{0}-\omega \right),
\label{rl}
\end{equation}
where 
 $\left|  \Psi_{0/f} \right>$ and $E_{0/f}$ are the wave functions
and the energies of  ground and final states, respectively.
\begin{figure}[htb!]
\centering
\epsfig{file=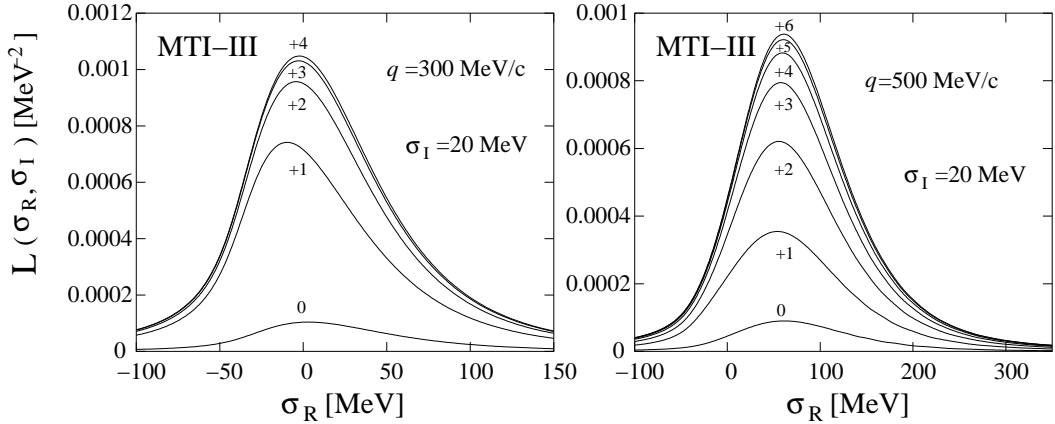, width=14cm}\\
\caption{ The LIT of the various isovector Coulomb multipoles,
  consecutively summed, as a function of the parameter $\sigma_R$ with
  $\sigma_I=20$ MeV fixed.}\label{fig1}
\end{figure}
The complication of the explicit calculation of all final states in
Eq.~(\ref{rl}) is circumvented via the LIT method, where one has to
solve the bound-state-like equation
\begin{equation}
(\hat{H}-E_0-\sigma_R+i \sigma_I)
 \left| \tilde{\Psi}  \right\rangle = 
\hat{\rho}({\bf q})  \left| \Psi_0 \right\rangle ~{\rm
  with}~L(\sigma_R,\sigma_I)=\left \langle \tilde{\Psi} | \tilde{\Psi} \right\rangle,
\label{eq2}
\end{equation} 
where $\sigma_R$ and $\sigma_I$ are the parameters of the transform $L$.
For the calculation we expand the charge operator into Coulomb
multipoles \cite{Eis70},
separating  them into isoscalar and
isovector parts.
The expansion is truncated when convergence is  achieved.
In Fig.~\ref{fig1} we show the LIT of  the isovector multipoles for momentum
transfers $q=300$  and  $500$ MeV/c: one readily  notes that for the lower
momentum transfer five multipoles are enough to reach convergence,
while for the higher momentum transfer two additional multipoles need to be considered.
In Fig.~\ref{fig2} we  present the result for the longitudinal 
response function for $q=300$ and  $500$ MeV/c, which is achieved  by inverting
the transform for each multipole.
 As in a previous calculation of $R_L$ with the  Trento (TN) potential
 \cite{el},  one can note that the semirealistic interaction leads to
 a rather good overall
description of the 
experimental data from Bates \cite{data_bates} and Saclay
\cite{data_saclay} for the
longitudinal response function.
The only deviation we observe is a pronounced peak close to threshold, which is due to
the monopole excitation of $^4$He. 
\begin{figure}[htb!]
\centering
\epsfig{file=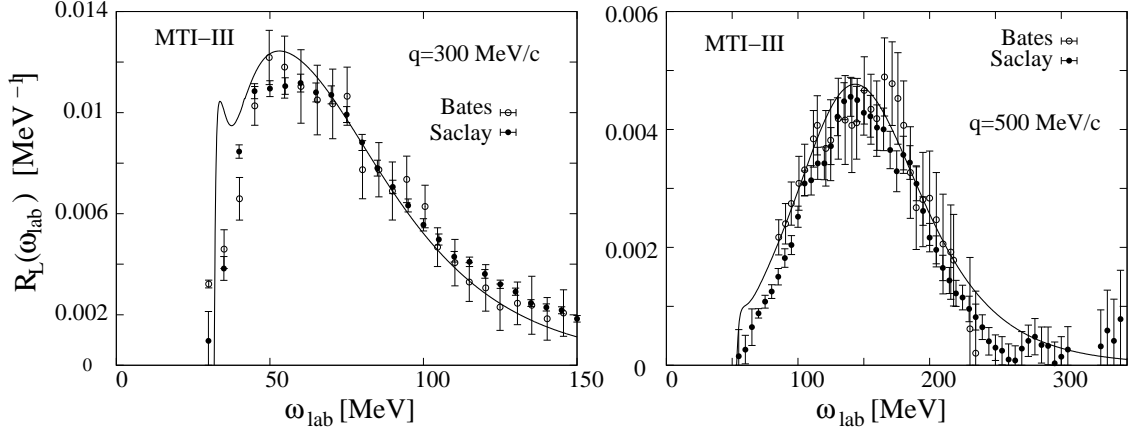, width=15cm}\\
\caption{$R_L$  with the  MTI-III potential as function of the
 laboratory energy for momentum transfers $q=300$ and  $500$ MeV/c.}
\label{fig2}
\end{figure}

\section{TRANSVERSE RESPONSE FUNCTION}

The transverse response function is defined as
\begin{equation}
R_T(\omega,{\bf q })=\int  \!\!\!\!\!\!\!
\sum_f\left| \left\langle \Psi _{f}\right|\hat{\bf
      J}_T({\bf q}) \left| \Psi _{0}\right\rangle \right| ^{2}\delta
\left(E_{f}-E_{0}-\omega \right),
\label{rt}
\end{equation}
where  ${\bf J}_T({\bf q}) $ is the  transverse electromagnetic
current operator.
The corresponding  bound-state-like equation  is the same as in
Eq.~(\ref{eq2}), where $\hat{\rho}({\bf q})$ is replaced by  $\hat{\bf
  J}_T({\bf q})$.
The  transverse current  includes one-body and two-body operators.
A two-body current is required in order  to
satisfy the continuity equation,
and has therefore to be consistent with the NN interaction used.
We derive a consistent MEC for the  MTI-III potential, which  is based on the
exchange of two  effective scalar mesons \cite{MT}.
The two-body current takes the form
\begin{equation}
{\bf J}_2 ({\bf q})=\frac{1}{4\pi^3}
e^{i{\bf R}\cdot {\bf q}}\big({\bf \nabla}_{\bf r}
I_m({\bf q},{\bf r})\big),
\end{equation}
where the function $I_m$ contains the meson propagator (for details see
Ref. \cite{mec}), ${\bf r}$ is the relative distance between the two
particles and ${\bf R}$ is the center of mass of the two-body sub-system. In
our calculation we neglect the ${\bf R}$ dependence 
for the sake of numerical simplicity, setting  $e^{i{\bf R}\cdot {\bf
    q}}\approx 1$. We therefore restrict
ourself to the case of low momentum transfer ${\bf q}$, where this
approximation is valid.
 \begin{figure}[htb!]
\centering
\epsfig{file=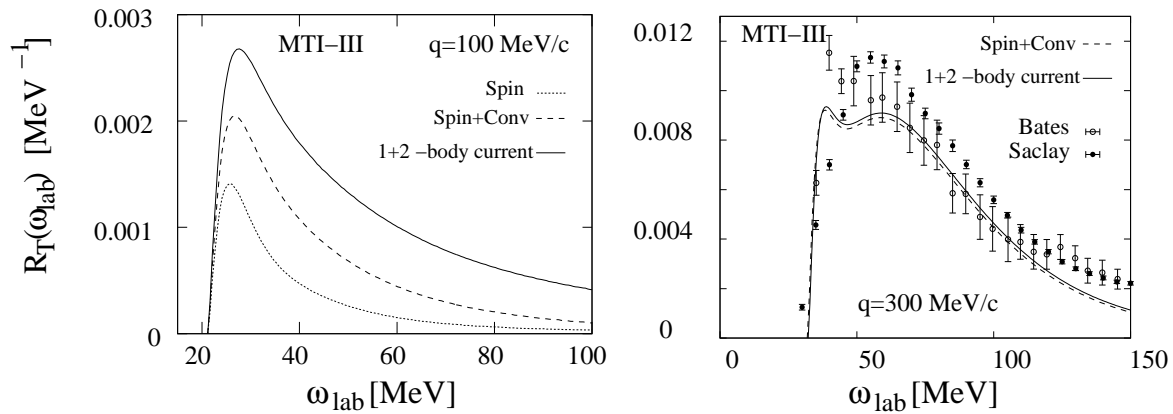, width=15.5cm}\\
\caption{Transverse response function: effect of one-body and two-body
  currents for $q=100$ and $300$ 
  MeV/c in comparison with the available experimental data from Bates
  \cite{data_bates} and Saclay \cite{data_saclay}.}
\label{fig3}
\end{figure}

In Fig.~\ref{fig3}  we show the transverse response function for the
different parts of the current operator.
The spin current  strongly dominates at the higher momentum
transfer $q=300$ MeV/c (therefore we do not show the convection
current contribution separately), while the effect of  the convection current
 is still important  at $ q= 100$ MeV.
One can see that  MEC plays an important role
 at $q=100$ MeV/c, but is very small  at $q=300$
MeV/c, about $2-3\%$ in the peak.
 At a  momentum transfer of  $q=300$ MeV/c  the additional
 contribution of the two-body current is not enough to describe
 satisfactorily the
experimental strength in the quasi-elastic peak.

\section{CONCLUSIONS}

We have presented the first calculation of the inclusive longitudinal and
transverse response functions of $^4$He within the LIT and EIHH methods.
Good agreement with available experimental data is found  for the longitudinal
response function, while some strength is still missing in the
transverse response function. Strong MEC effect are found at low
momentum transfer, where unfortunately no experimental data are
available. Within our semirealistic potential model
and consistent MEC  we do not find a strong two-body current
effect  at $q=300$ MeV/c  as obtained in Ref.~\cite{Carl_Schiav}. This
is probably due to the missing
explicit pionic degrees of freedom in our model.

\end{document}